\documentclass[a4paper,11pt]{article}
\pdfoutput=1 

\usepackage{jinstpub} 
\usepackage{comment}

\title{\boldmath Impact of the positive ion current on large size neutrino detectors and delayed photon emission}

\author[a,1]{R.~Santorelli,\note{Corresponding author.}}
\author[b]{S.~di~Luise,} 
\author[a]{E.~Sanchez Garcia,}
\author[a]{P.~Garcia~Abia,}
\author[b]{T.~Lux,} 
\author[a]{V.~Pesudo,}
\author[a]{L.~Romero}

\affiliation[a]{Centro de Investigaciones Energ\'{e}ticas, Medioambientales y Tecnol\'{o}gicas (CIEMAT), Av. Complutense 40, 28040 Madrid, Spain}
\affiliation[b]{Institut de F\'{i}sica d\'{}Altes Energies (IFAE), Campus UAB, Bellaterra (Barcelona), Spain}

\emailAdd{roberto.santorelli@ciemat.es}

\abstract{Given their small mobility coefficient in liquid argon with respect to the electrons, the ions spend a considerably longer time in the active volume. We studied the effects of the positive ion current in a liquid argon time projection chamber, in the context of massive argon experiments for neutrino physics. The constant recombination between free ions and electrons produces a quenching of the charge signal and a constant emission of photons, uncorrelated in time and space to the physical interactions. The predictions evidence some potential concerns for multi-ton argon detectors, particularly when operated on surface.}

\keywords{Only keywords from JINST's keywords list please}

\arxivnumber{1234.56789} 

\proceeding{LIDINE 2017: Light Detection in Noble Elements\\
22-24 September 2017\\
SLAC National Accelerator Laboratory}

\begin{document}
\maketitle
\flushbottom

\section{Introduction}
\label{sec:intro}
Both photons at 128 nm and ion/electron pairs are generated simultaneously by particle interactions in  argon.  In a typical liquid argon time projection chamber (LAr-TPC),  photon sensors are used to detect the  prompt scintillation light, while the electrons drift to the anode  by mean of a constant electric field $\vec{E_{d}}$. The drift field prevents the full recombination of the opposite charges created in the ionization track, reducing the so-called recombination light and making possible the collection of the ionization signal. In case of a single-phase TPC, the electrons are directly collected by thin wires placed  in the liquid \cite{Amerio:2004ze}, while, in case of a double-phase liquid/vapor chamber, they are extracted to a gas region placed above the sensitive volume \cite{WA105:technical}. In the latter case, the charge can be amplified  through a sufficiently intense electric field ($\approx$ 30 kV/cm) with the production of a Townsend avalanche with electron/ions pairs.

The positive and negative charges created in the LAr bulk drift along the same  $\vec{E_{d}}$ field lines, although the ions have a mobility,  thus a velocity, much smaller than the electrons ($v_{i}  \ll v_{e}$). Considering a typical field, $E_d = 1$~kV/cm, the electron velocity is of the order of $v_e \approx 2$~mm/$\mu$s \cite{Walkowiak:2000wf}. The ion mobility is not very well know, and the values reported in the literature range between $\mu_i~\approx~2\cdot~10^{-4}~ \text{cm}^2\,\text{V}^{-1}\text{s}^{-1}$ \cite{Dey:1968} and $\mu_i \approx 1.6\cdot 10^{-3}~\text{cm}^2\,\text{V}^{-1}\text{s}^{-1}$ \cite{ICARUS:2015torti} with the liquid in steady state. Even considering the more conservative larger value, the expected ion velocity at 1~kV/cm is $v_i \approx 1.6\cdot 10^{-5}$~mm/$\mu$s, which is five orders of magnitude lower than that of the electrons at the same field.

Once a drift field is applied, the electrons are promptly collected while the ions stay in the LAr for much longer time. As a consequence, the positive charge density can increase with time, reaching a constant  value dependent on the total ionization produced and on the amplitude of the  field itself. As a consequence, in stable conditions, the average density of positive ions is much larger than that of electrons ($d_{I}  \gg d_{e}$). 

This space charge can locally modify the drift lines, the amplitude of the electric field, and ultimately the velocity of the  electrons, thus, a  displacement in the reconstructed position of the ionization signal can be produced. Additionally, for relatively large values of $d_{I}$, the positive-charge density can be sizable such that the probability of a ``secondary electron/ion recombination'', different than the recombination that occurs within the ionization track \cite{Chepel:2012sj}, has to be considered between the drifting  electrons and the free positive ions. The effect can cause an additional signal loss, with a probability dependent on the electron drift path, that could resemble the charge quenching given by the contamination of the electronegative impurities in LAr.  At the same time it can produce the emission of photons through channels similar to the recombination light.  

This effect can be particularly relevant for double phase detectors foreseeing large charge amplification factors, where the ions, created in the vapor volume, may cross the gas-liquid interface  and further increase  $d_{I}$ in the active LAr volume.

\section{Charge density in steady state}
\label{sec:FE}

In an underground LAr experiment, the dominant contribution to the charge production is typically given by the  $^{39}$Ar decay, a $\beta$ emitter with a Q-value of 565~keV whose activity in natural argon  is $\approx$~1~Bq/kg \cite{Benetti:2006az}, or, for liquid argon in standard conditions, $\approx1400$~Bq/m$^3$. Assuming that the mean energy deposited  per decay in the active volume is approximately one third of the total Q-value, and considering that the mean energy required to create an ion-electron pair in LAr is W = 23.6~eV \cite{Miyajima:1974}, one $^{39}$Ar decay produces in average $\approx8\cdot10^3$~pairs, therefore the  ionization rate, $h_0$, due to $^{39}$Ar is $\approx1.1\cdot10^7$~pairs/(m$^3$s).

In case the detector is located on the Earth surface, the contribution to the total ionization produced in the active volume is mainly given by the muons, with an average flux at sea level of $168~\text{muons}/(\text{m}^2\text{s})$ \cite{Grieder:2001ct}. Considering that most of the muons are at their minimum ionizing energy, the energy loss in argon as a function of the density is
$\approx1.5~\text{MeV}/\text{cm}^2/\text{g}$, and the average energy deposited is 210~MeV/m per muon or $35$~GeV/(m$^3$s) in liquid argon. Assuming the W value above, the ion production rate is  $h_0=1.5\cdot 10^9$ pairs/(m$^3$s), two orders of magnitude greater than that of the $^{39}$Ar decay.

\begin{figure}
 \begin{center}
  \includegraphics[height=.41\textheight]{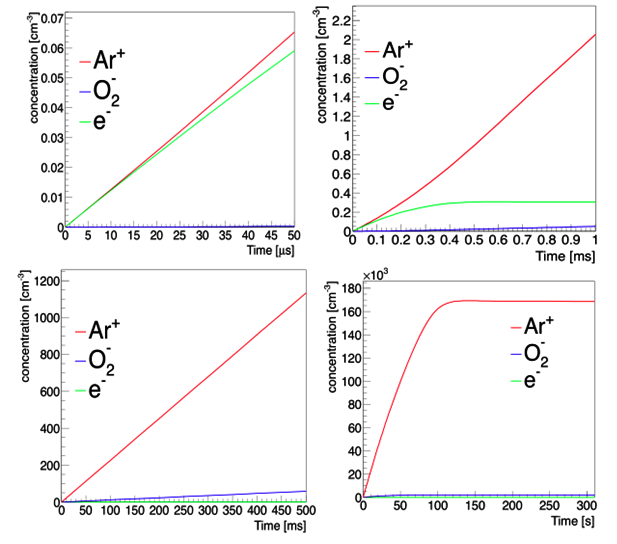}
  \caption{Evolution with time of the different density of species in LAr in four timescales (increasing from left to right, and from top to bottom). After a few seconds the positive-ion density is several orders of magnitude larger than that of electrons and negative ions.}
  \label{COM_UG}
  \end{center}
\end{figure}

At zero field, the opposite charges created in liquid promptly recombine producing photons. Once the electric field is present, the  \textcolor{black}{columnar} recombination of the electrons with the parent ions is suppressed, and the opposite charges drift toward the anode and the cathode. The average number of free charges escaping the recombination, $h$, is a function of the drift field, and is usually approximated with the so-called Birks  law, $h={h_0}/(1+k_E/ E_d)$ \cite{Birks:1964}. The measured value of constant $k_E$ in argon is $0.53 \pm 0.04$~kV/cm~\cite{Scalettar:1982}.

Given the much larger velocity of the electrons with respect to the ions, the negative charge created is promptly extracted from the LAr volume in a timescale of ms ($v_e \approx 2$~m/ms  with $E_d \approx 1$~kV/cm). At the same time, minutes are necessary for the ions to drift entirely along the same volume ($v_i \approx 1.6\cdot10^{-2}$~m/s), thus a positive charge density, much larger than the negative one,  builds up in the LAr bulk. Electronegative molecules, like oxygen and water, diluted in the argon volume may trap the drifting electrons, creating negative ions which have a drift velocity similar to the argon ones. The negative-ion density can partially cancel the effects of the positive space charge.
A stationary condition is reached when all the concentrations do not change significantly anymore with time. Both timescale of the species evolution and steady value depend on the drift field, ionization rate and impurities concentration.

The solution to this problem is calculated with a Finite Element Analysis (FEA) approach. A model is implemented in COMSOL Multiphysics~\cite{COMSOL}, including the ``Electrostatics'' and ``Transport of Diluted Species'' modules. 
The problem consists of calculating evolution and stationary state of the ions and electrons in a $1\times 1 \times 1$ m$^{3}$ box filled with liquid argon, considering the ionization produced mainly by cosmic muons and a 1 kV/cm drift field. Diffusion of the charges and recombination is taken into account. 

The evolution of the species on four different timescales is shown in Fig.~\ref{COM_UG}. Given their relatively fast velocity compared to the other species,  the free electrons  reach a constant density in less than one ms, with an average $d_e$ value of the order of 0.3 cm$^{-3}$.  This value are, as expected, comparable with the maximum drift time for the  considered geometry. Both the negative and positive ions reach a steady state within a timescale of the order of one minute. The amount of electronegative impurities is much larger than the amount of drifting
electrons, however, assuming an electron lifetime in LAr of the order of 5 ms or more, the rate of formation of negative ions is much smaller compared to the ionization produced. The average negative-charge density (both electrons and ions) is negligible  compared to the positive one, which is about $1.7\cdot10^{5}$ cm$^{-3}$ or d$_I\approx1.7\cdot10^{11}$ m$^{-3}$. A similar value can be derived from the analytic estimation reported in \cite{Romero:2016tla}. The results demonstrate that, in steady state, a massive space charge is present in the LAr bulk, thus its impact on the drift field and charge collection has to be evaluated.

\section{Field distortion and secondary electron-ion recombination}
\label{sec:FL}

\begin{figure}[t!]
\begin{center}
  \includegraphics[height=.28\textheight]{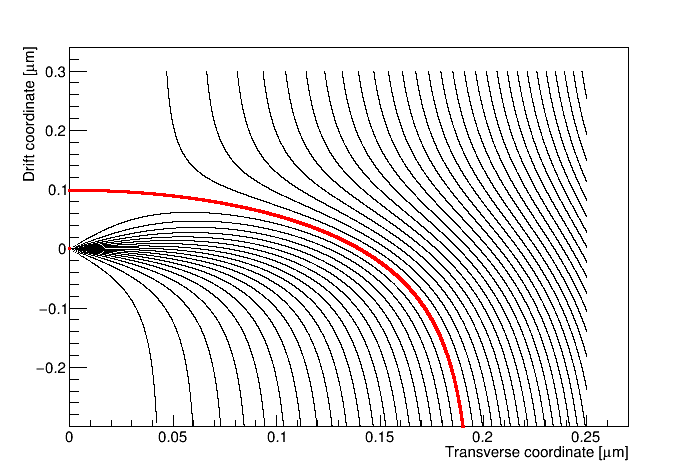}
  \caption{Configuration of the drift lines near a positive ion placed at (0, 0) for a nominal drift field of 1~kV/cm. The thick red line is the envelope of the field lines ending on the ion. }
  \label{Field_lines}
  \end{center}
\end{figure}

The positive-charge density  can modify the effective electric field. In the limit of a null ion current, the drift field is constant and it is equal to the cathode voltage divided by the total drift length,~$L$. On the contrary, an ion cloud makes  the  field to change with the drift coordinate $l$, from a minimum at the anode to a maximum at the cathode. In order to achieve the desired minimum field value, it is necessary to increase the nominal cathode voltage by a factor dependent on the charge density.
At the same time the space charge modifies the direction drift lines and the velocity of the electrons  produced in the liquid, leading to a  displacement in the reconstructed position of the ionization signal.  Fig.~\ref{Field_lines} shows the field paths approaching an ion positioned at (0, 0), which has a negligible size at the $\mu$m scale, for $E_d = 1$~kV/cm. The red curve is the envelope of all the field lines ending on the ion (detailed calculation in~\cite{Romero:2016tla}). The ion is motionless compared to  the free electrons drifting by means of the electric field, which is not uniform given the presence of the ion itself. The field lines bend and, depending on the distance between  the charges, a displacement in the transverse coordinate is produced when the electron approaches the ion. The lines in a circle of radius  $\approx0.2~\mu$m end on the ion, thus a free electron drifting along those lines recombines and disappears. 

We define the cross section $S_{CS}$ as the transverse area whose crossing field lines end on one ion. The section should be far enough from the ion such that the ion field is negligible compared to the drift field, and all the lines emerging from the ion cross that section. The total number of field lines emerging from the ion, $q/\epsilon$, is equal to the number of lines traversing the cross section, $E_{d}\cdot S_{CS}$, therefore $S_{CS}=q/(\epsilon E_{d})$, where $q$ is the elementary charge, $\epsilon$ the absolute permittivity of the liquid argon and $E_d$ the amplitude of the drift field. For a typical $E_d$ value of  1~kV/cm,  $S_{CS} = 1.2\cdot10^{-1} \mu$m$^2$ ( $S_{CS}=1.2\cdot 10^{-13}~$m$^2$) is much larger than the ion dimensions. The total cross section, obtained considering all the ions constantly present within the drift volume can be macroscopic and  the effect of the recombination cannot be neglected.

\begin{figure}[t!]
\begin{center}
\includegraphics[height=.29\textheight]{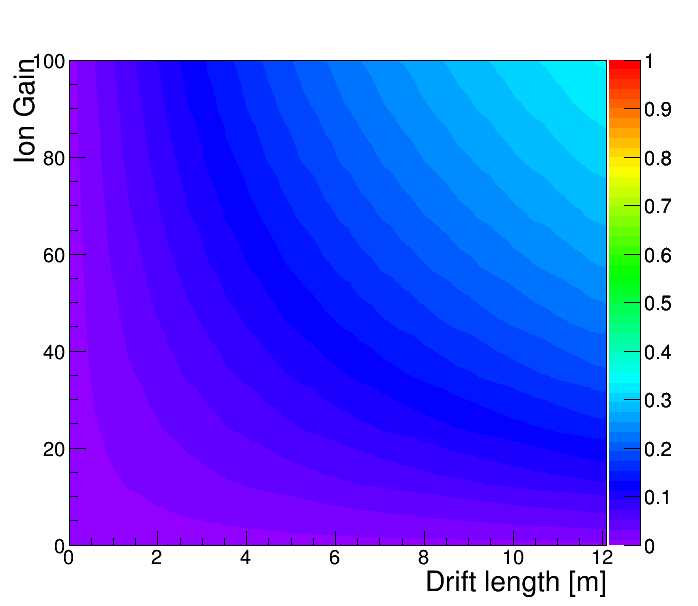}	
  \includegraphics[height=.29\textheight]{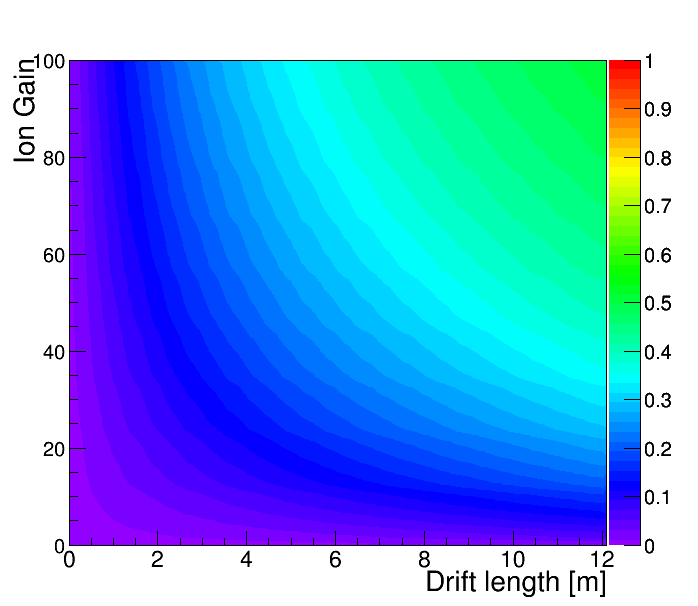}
  \caption{Expected recombination probability (color scale) as a function of the ion gain $G_I$ and drift length in case of a detector like the DUNE double-phase module (L = 12 m and charge amplification), considering $E_d$ = 1.0 kV/cm ($left$) and $E_d$ = 0.5 kV/cm ($right$).}
 \label{Losses_UG_Dune}
  \end{center}
\end{figure}

As discussed in sec. \ref{sec:FE}, the expected positive-ion density in steady conditions for a shallow detector is $d_{{I}}~\approx~1.7\cdot10^{11}$m$^{-3}$, thus the secondary recombination can be significant taking into account a single-phase detector with few meters drifting length.  The effect is expected to be even more relevant in case of a dual-phase detector due to ion feedback from the gas, which can increase the $d_{{I}}$ value further. We introduce the ion gain, $G_I$, defined as the number of positive ions  injected into the liquid  for each electron extracted. This factor is proportional to the  electron  amplification $G$ through a constant $\beta$ ($\beta \leq 1$)  which takes into account the average loss of the positive charge in the gas, given by the ions scattering onto field lines not ending on the liquid surface, as well as the efficiency to pass the liquid-gas interface. If the amplification factor $G_I$ is large enough, the positive-charge density, $\rho_i$, in the liquid can be widely increased by the secondary ions produced by the Townsend avalanche, thus the secondary recombination can be relevant also in case of deep underground detector whose ion density is only given by the $^{39}$Ar decay. The recombination probability calculated for a detector like the DUNE double-phase module~\cite{Acciarri:2016crz} is displayed in Fig.~\ref{Losses_UG_Dune}, considering 1~kV/cm and 0.5~kV/cm effective drift fields. In addition to the charge quenching, a light emission is expected through a mechanism similar to the columnar recombination. The effect can reduce significantly the electron signal increasing the scintillation light, although on timescales which, depending on the drift length and electron velocity, could be of the order of
several ms. Depending on the interaction rate in liquid and the detector photon detection
efficiency, a constant emission of pulses from the liquid argon bulk could be detected. At 0.5 kV/cm up to  50~\% of the electrons created near the cathode are expected to produce a recombination photon along the drift. 

It is important for a double-phase massive detector with drift of many meters and charge amplification to investigate the ion feedback from the gas, possibly measuring $G_I$. An investigation of this topic is currently carried-out by the CIEMAT and IFAE groups working on the LAr technology.

\section{Conclusions}
Given the much larger velocity of the electrons with respect to the ions, a massive positive-charge density is created in a LAr detector. A finite element analysis considering a detector placed on the Earth surface, whose most important ionization source is yielded by muons, shows that, in steady state, both the electron and negative-ions density are negligible compared to the density of positive ions. This density can be further increased in case of ion feedback from the gas. A relevant recombination probability of the free charges is expected along the drift field lines, quenching the electron signal in large detectors and producing a constant light emission in the LAr bulk. This emission is poorly correlated in space and time to the particle interaction and it depends on the electric drift field and event rate inside the detector.


\acknowledgments
This research is funded by the Spanish Ministry of Economy and Competitiveness (MINECO) through the grant FPA2015-70657P. The authors are also supported by the ``Unidad de Excelencia Mar\'{i}a de Maeztu: CIEMAT - F\'{i}sica de Part\'{i}culas'' through the grant MDM-2015-0509.


\end{document}